# NEWSPAPER STORY PROBLEMS AND OTHER TASKS FOR CONTEXT BASED PHYSICS EDUCATION:
# A RESEARCH BASED REPORT ON CLASSROOM PRACTICE

Kuhn, Jochen[1]; Müller, Andreas[2]; Vogt, Patrik
[1] Ludwig-Maximilians-Universität München / LMU Munich
[2] University of Geneva (andreas.mueller@unige.ch)

## ABSTRACT

*Background:* Context Based Science Education (CBSE), that is using authentic, real-life contexts, has a long-standing tradition and is discussed as a highly promising approach in science education. In particular, it is supposed that CBSE can foster pupils' engagement for working on, and learning with, properly chosen context-based problems. However, classroom implementations of CBSE based on solid empirical evidence are surprisingly scarce.

*Purpose*: The present research-based report of practice seeks to bridge this theory practice gap for some specific forms of CBSE.

*Sample/Setting*: We examine the use of science problems based on newspaper articles and the real-life contexts they provide ("newspaper story problems" or NSP). A broad sample (N > 1,600) of secondary level I pupils of two age groups and various academic backgrounds in Germany was studied.

*Design and Methods*: While the research background (methodology, statistical analyses, results) has been reported elsewhere (Kuhn, 2010; Kuhn & Müller, 2014), the main objective of the present contribution is to provide a detailed account of the practical aspects of the approach (examples, classroom implementation, etc.).
Two concrete, curriculum-relevant classroom teaching experiments based on newspaper story problems are reported, combined with a quasi-experimental study comparing NSPs against conventional textbook problems. They covered two topics and two related age groups (elementary kinematics, 7th/8th grade; energy, 9th/10th grade). The implementation of the teaching-learning sequence in classroom practice is described in detail, especially the design principles of "embedded data", "adjustable problem complexity", and "active and autonomous problem solving". For the empirical investigation, research-validated instruments for various motivational and learning outcomes were used, in particular for perceived authenticity/link to real life and for transfer. Thorough measures to ensure comparability (e.g., same lesson plan and teacher) were implemented. The entire development and research project was carried out in a research-practice co-operation network for physics education, including more than 20 teachers from six different types of school. Additionally, tasks types similar to NSPs, but using other ways of contextualisation (e.g. advertisements) will be discussed as perspective.

*Results*: A considerable improvement in motivation was found, which proved stable at least in the medium term. Learning turned out to also be fostered to a sizeable extent, including the educationally important issue of transfer. Moreover, these positive effects were "robust" in the sense of not being affected (or only weakly so) by potentially relevant pupil characteristics (such as gender, or language-, maths- and physics-related academic ability).

*Conclusions*: The use of NSPs as a form of CBSE can have large positive, robust, and sustainable effects of both motivation and learning. Being flexible and practical to implement, they appear thus highly suited to classroom application. In perspective, a series of similar forms of tasks implementing CBSE is presented, such as by experimental of aesthetic contexts.

## 1. INTRODUCTION

The present work starts off from a set of closely related, widespread problems in science education.[1] First, students' transfer of knowledge, especially to real-life contexts, is low. Indeed, even when knowledge in some initial learning context is successfully acquired, students fail at application and transfer problems outside said context. Such "inert knowledge" has long been identified as a central problem of science education and educational psychology (Renkl 1996; Renkl et al., 1996; Whitehead, 1929). Second, sense-making and knowledgeable usage of science-related texts has been found wanting. These two problems have recently been confirmed, for example, by large-scale assessment studies. This has been particularly striking in the case of students in Germany (Baumert et al., 2001; 2002). Third, these problems are embedded within the general problem of students' low motivation for and (sometimes even basic) understanding of science in many countries. Strong evidence of this is found in the ROSE (where the focus is on attitude/motivation; Sjoberg & Schreiber, 2010) and PISA (OECD 2010, Fig. I.3.19) studies. It is here that Context-Based Science Education (CBSE) comes into play. CBSE has a long history and is advocated as a highly promising approach, particularly for physics education (Duit & Mikelskis-Seifert, 2007; Taasoobshirazi & Carr, 2008; Kuhn et al., 2010a) and science education in general (Fensham, 2009; Bennett et al., 2007; Parchmann et al., 2006; Nentwig & Waddington, 2005). In the current broad understanding, CBSE is defined as "using concepts and process skills in real-life contexts that are relevant to students from diverse backgrounds" (Glynn & Koballa 2005, p.75). Making science issues relevant to students and their social groups – their families and peers – or trying to do so counters the wide-spread perception of physics (especially) as being dry, impersonal, and irrelevant, and this is supposed to have positive effects both on motivation and learning. Moreover, CBSE should be targeted at students with "diverse backgrounds" (i.e., with a broad target groups), and its (supposed) benefits should not be restricted to a select student population.

---
[1] While this specific study was on physics education, many of its statements and findings also hold for science education in general.

The notion of "authenticity" – in the sense of a close relationship to real (or at least realistic) contexts ensuring "relevance to students' interests and lives" – is also central to PISA's understanding of scientific literacy (OECD, 2007, p.36). Importantly here, such tasks are usually not formulated in mathematical or scientific terms, they refer to real-world events or objects, in common, every-day language. Thus, a work of "translation" with terminological and conceptual reframing must be carried out, representing a first step of cognitive activation (OECD, 2006; Weiss & Müller, 2015).

Drawing on past research as well as on good practice reports from mathematics, physics, and other science classrooms (Jarman & McClune, 2007), the present contribution deals with some specific ways of establishing such contexts, using science problems based on newspaper articles and the real-life contexts they provide (hereafter "newspaper story problems" or NSP), see Fig. 1. With the general goal being to bridge physics education research and classroom practice for the issues under consideration, the following objectives will be addressed:

- to provide empirical evidence for NSPs' reported potential to improve both motivation and learning, including transfer, and thereby closing the existing research-practice gap (see abstract and sect. 2) for a specific example of context-based physics education;
- to investigate whether such benefits, should they exist, can in fact be established for more than a short-term duration, and for pupils[2] from diverse backgrounds, and in different teaching and school settings (for the latter aspect, key features of a research-practice co-operation network for physics education involving a total sample of about 1,600 pupils will be described);
- to give, of course, examples and more generally, principles of classroom implementation for NSP, enabling interested teachers to adapt them for their own use; the present paper will mainly deal with NSP (on the topics of elementary kinematics and energy (transformations) with a brief outlook on similar context-based science learning activities.

While the research background (methodology, statistical analyses, results) has been reported elsewhere (Kuhn, 2010; Kuhn & Müller, 2014), the main objective of the present contribution is to provide a detailed account of the practical aspects of the approach (examples, classroom implementation, etc.).

## 2. NEWSPAPER STORY PROBLEMS: BAKGROUND AND RATIONALE

Evidence in favour of CBSE has arisen from both research and classroom practice. However, other results in science education show that taking physics contexts into account in the sense of "making it relevant" (Nentwig & Waddington, 2005) is not sufficient to maintain (nor generate) pupil interest. Figure 2 shows the development of interest in and perceived relevance of physics from 5th to 10th grade, that is, from beginning to end of secondary level I in Germany (Muckenfuß, 1996). A striking "scissor-shaped" pattern is visible: while perceived relevance of physics increases over the years considered, interest shows the marked drop we are all aware of and face in our daily teaching. This strong divergence of ascribed relevance and personal interest has been confirmed by the large-scale international study, ROSE (Sjoberg & Schreiber, 2010). This finding can be interpreted in that from the end of childhood (5th grade) to young adulthood (10th grade), youths become increasingly aware of the relevance of physics (science) for the world they live in, but that does not entail personal relevance or interest. The latter being absent, one sees a decline of interest that is well established in developmental psychology studies for this age group in science and almost all other school subjects as well (Häussler & Hoffmann, 1995; CHSN, 2009; Krapp & Prenzel, 2011).

Given this, at least two questions must be answered: First, what contexts to choose, and second, how to embed CBSE into classroom practice, so that it is in fact perceived as interesting (and cognitively activating) by the pupils themselves. We will refer to these two questions in turn as kind of context, and classroom implementation, respectively.

It is for the question raised by Fig. 2 that a specific form of establishing contexts in the classroom comes into play, namely, "stories as context". Embedding learning content in an interesting story is current and good practice in many science (and non-science) classrooms. One way of doing this is through newspaper

---

[2] "Pupil" is still currently used in British English (e.g., according to the *Cambridge Dictionary*: "a child at school"), and as it correctly differentiates the institution (and age group) it is preferred to the undifferentiated American English "student".

| a) "Newspaper story problem" format | b) "Conventional task" format |
|---|---|
| **Lausanne/ngn.** Another great round-a-world adventure a la Steve Fossett's solo flight last month. Now it seems the explorer Bertrand Piccard will attempt the world's first solar powered around the world flight.<br>Piccard comes from a family of explorers and made history in March of 1999 with a nonstop, around-the-world flight in a hot-air balloon, the Breitling Orbiter 3. But what about the solar power side of this? Is that really possible? Nearly the entire body of the plane will be covered by 287 square yards (240 square metres) of solar panels. Piccard estimates that enough power can be generated to sustain a flight of roughly 60 miles an hour (97 kilometres an hour). The plane's batteries are going to have to be pretty heavy, capable of storing 200 watts per kilogram, so that the plane can run at night.<br>*Gadling*, 2005-04-13 | In 2007 the explorer Bertrand Piccard will attempt the world's first solar powered around the world flight. He estimates that the solar panels of the plane can generated enough power to sustain a flight of roughly 60 miles an hour (97 kilometres an hour). The plane's batteries are going to have to be pretty heavy, capable of storing 200 watts per kilogram, so that the plane can run at night. |

**c) Tasks common to both formats**

1. How long will Bertrand Piccard be on his way around the world?
2. How much electrical energy per kilogram and how much power per kilogram has to be produced at least?
3. How much electrical energy per kilogram and how much power per kilogram has to be produced at least by the solar panels, if Piccard will drive ca. 75% of the flight by day?
4. How much electrical energy per kilogram has to be produced at least by the batteries?
5. Discuss your results critically, e.g. with respect to the energy transformation process, and use physical arguments thereto.

**Fig. 1.** A newspaper story (a) and conventional (b) problem on the topic of electrical energy. Tasks are identical in both cases (c).

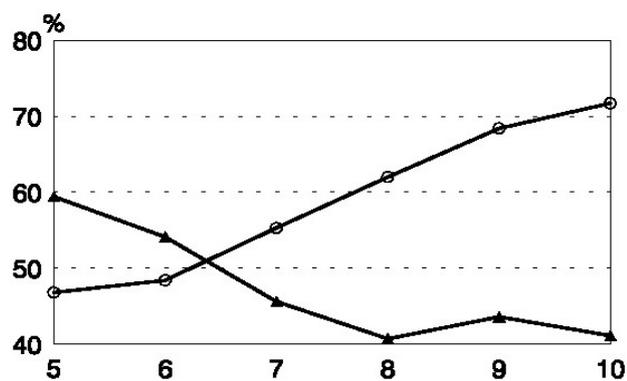

**Fig. 2.** Development of interest (▲) in and perceived relevance (o) of physics from beginning to end of secondary level I in Germany; *x*-axis: 5th to 10th grade, *y*-axis: percentage of maximum possible (POMP; Cohen et al., 1999) of the scale in question (Muckenfuß, 1996)

Given this, at least two questions must be answered: First, what contexts to choose, and second, how to embed CBSE into classroom practice, so that it is in fact perceived as interesting (and cognitively activating) by the pupils themselves. We will refer to these two questions in turn as kind of context, and classroom implementation, respectively.

It is for the question raised by Fig. 2 that a specific form of establishing contexts in the classroom comes into play, namely, "stories as context". Embedding learning content in an interesting story is current and good practice in many science (and non-science) classrooms. One way of doing this is through newspaper story problems. These are based on newspaper articles covering science related issues and ask pupils to work on various related questions and tasks.

Newspaper story problems thus include two components: (i) an article (or excerpt) taken from a real newspaper and (ii) one or more tasks emerging from this text (see Fig. 1). To maintain a "flavour" of authenticity and connection to the real world, these articles are used with little or no modification (give-or-take possible editing for length). The rationale behind the use of NSPs is, of course, that newspaper articles (almost by definition) provide out-of-school, real-life contexts. Moreover, journalists are supposed to be experts at writing interesting, high-quality stories (and it is a good thing to use their know-how). An excellent introduction and source of ideas on using newspapers to enrich science teaching is offered by Jarman and McClune (2007). For various specific disciplines, reports of good practice and sample collections for the use of newspapers exist (extensively in mathematics, (Paulos, 1995; Herget & Scholz 1998), and to some extent also for physics (Armbrust, 2001) and chemistry (Toby, 1997; Haupt, 2005). Jarman and McClune (2002) review the use and objectives of teaching science with newspapers. Note that in their study "links with everyday life" were by far those most frequently stated by teachers as their main intention for (76%) and the main benefit of (62%) using newspapers in the classroom. Establishing contexts through the educational use of newspapers also has a long tradition for general literacy purposes, and it is at the heart of current "newspapers in education" programmes run by several national newspaper associations (KMK/BDZV, 2006; NAAF, 2007; 2010). Thus, a question of the present study was to investigate whether newspapers would provide a kind of context pupils would perceive not only as relevant, but also as interesting.

As explained above, together with the choice of a kind of context for teaching, NSP in the present study, a decisive question is that of classroom implementation (i.e., how to embed CBSE into classroom practice). Bearing in mind the general problem of low student motivation for physics and the more specific problem of the "Muckenfuß scissor" (Fig. 2), a well-founded choice for an effective mode of implementation is just as important as it is for the kind of context used. The mode of implementation must also be compatible with basic choice of NSP in the first place.

The research literature offers a quite promising solution for this challenge, namely an approach called "anchored instruction" (AI). The starting point of AI is the idea that it is important to anchor teaching and learning in the most authentic contexts that require learners to solve meaningful problems. The crucial point of this approach is building of cognitive structures as important element for learning. It depends on the 'anchor ground' for anchoring of new learning material. Situatedness is achieved in AI by providing a media context, which serves as an "anchor of interest" and is the starting point for the development of problem-solving. This is an internationally renowned educational approach, with its central idea of providing contextualisation (or "situatedness") by means of "anchor media" (or "anchors", for short), that is, authentic, affectively, and cognitively engaging stories (with science or other curriculum related content). They are technically provided in the form of multimedia video discs (CTGV 1990, 1991), offering a 15-to-20-minute video sequence with real characters in a real environment. The sequence contains a complex problem to be solved by pupils, largely through independent work in small groups. While solving the problem, pupils can access passages of the sequence or background information available on the videodiscs. These multimedia video films (the "anchor media") make up the central element of AI, which in short can be characterised as a complex multimedia story problem. Ideally, the anchor is interesting enough to guide and challenge motivated work on the problems. Students are asked to put themselves in the shoes of the main character in the short film and solve the problem presented. Exploratory learning is embedded through the narrative structure of the presentations. Students are to experience insights and actively acquire the necessary problem-solving skills through their own actions. Thus, the learned knowledge is to be built flexibly beyond the school setting. The learners should recognize the meaningfulness, significance and relevance of the learning content, so that the solution of the problem is not only recognized as important for a good performance record or the achievement of the class goal. The development of anchor media should be guided by its design principles in particular.

Within the framework of a large, long-lasting research and development programme, AI arrived at a series of empirically tested design principles about how pupils should work and learn with these anchor media (CTGV, 1991).

Yet despite its thorough research foundation and considerable success in application to date, AI has some shortcomings. Creating such anchor media requires considerable resources (both practical and financial) beyond what is usually available in schools. Moreover, using the material requires substantial technical prerequisites (videodisc/multimedia) for all learning groups. Finally, it cannot easily be adapted to the demands and needs of a specific learning group or situation. The central idea of the present work is to combine the authenticity and "story character" of the original AI approach and existing research-based guidance on its implementation in the classroom on the one hand, with the practicability and flexibility offered by the easy-to-obtain and easy-to-deal-with medium of newspapers (replacing videodiscs) on the other. In contrast to multimedia anchors, authentic text (including images) can be obtained and modified with considerably less effort, in a reasonable amount of time, and with only moderate technical and material resources. Furthermore, this approach features the necessary flexibility in terms educational or technical conditions is relatively easy.

**Tab. 1.** Design features for authentic learning media within the original (AI) and modified (MAI) Anchored Instruction frameworks. These features were originally discussed in AI (CTGV, 1990; 1991), with the exception for the last two, which are included for the purpose of the present study.

| Design principle | AI | MAI/NSP |
|---|---|---|
| Affective anchor media, authenticity | Yes | Yes, empirically validated (cf. below and Fig. 6a and 7a) |
| Story character, "narrative" format | Yes, by videodisc | Yes, by newspaper article |
| Embedded data | Yes | Yes |
| Links across the curriculum (horizontal and vertical) | Yes | Horizontal: yes, by construction (NSP involving links to various other issues, such as societal, technological, biological, etc.). Vertical: yes, in principle (established in this study within a given problem, but possible also across problems in a more long-lasting implementation of the approach, covering several topics). |
| Related problems (by pairs or multiple) | Yes | Yes |
| Complex problems | Yes | Yes, in the sense of encompassing PISA transfer levels, cf. text (but not compared to AI); in general, adjustable complexity |
| Active and autonomous problem solving | Yes | Yes, according to problem-based learning cycle, cf. Fig 4 |
| Macro vs. micro approach | Macro | Intermediate, adjustable extent (cf. text) |
| Adjustability: content, complexity, extent | No (hardly), due to practical and technical difficulties, cf. text | Yes, cf. above and text |

It is for this reason that we developed and investigated an approach by transferring the design principles and construction features of AI to other anchor media (Kuhn & Müller, 2005), in particular newspaper story problems, which are the focus of the present paper. Given its point of departure, we will sometimes refer to this approach as modified anchored instruction (MAI).[3]

Current work on learning tasks with contexts in a more narrow sense typically addresses problem-solving tasks that have to be solved by applying subject-specific models. There is little doubt about the positive effects of such context-based approaches on affective variables (e.g., interest). At the same time, the observed variance in performance has so far been poorly explained. Various approaches have attempted to fill this research gap; for example, by examining the interplay of task characteristics such as authenticity or relevance and their interaction with person characteristics. However, an explanation for the heterogeneous findings on the influence of context on performance in the problem-solving process is still missing. In this context, performance measurement in problem solving usually takes place at the level of the entire process. Therefore, the influence of context can also only be studied in relation to this overarching measurement. In this context, recent work shows that the effect of context on performance in the problem-solving process varies considerably (Löffler & Kauertz, 2021). The positive effects on problem understanding and reasoned problem identification are about equal in strength to the negative effects on problem solving itself, and therefore cancel each other out at the level of the entire process. This gives rise to the demand to examine the process in detail in future context research.

We now turn to a detailed description of the underlying AI design principles and their application to NSP before reporting the effects of this approach on physics motivation and learning.

## 3. CLASSROOM IMPLEMENTATION: DESIGN PRINCIPLES

For context-based teaching and learning to be effective, its classroom implementation must meet certain basic criteria. As mentioned above, the AI approach features a series of research-based design principles. These are summarised in Tab. 1 and the most important (for NSP) will now be discussed in detail. In addition to that we add guidelines for creating newspaper articles for teachers following the NSP design principles in Tab. 2.

of situational or personal characteristics, an obvious condition for classroom implementation (regarding e. g., content, length, degree of difficulty, degree of guidance, to mention but a few important criteria). Thus, "tailoring" such anchors to the changing demands of the

### 3.1 Embedded Data

The basic data necessary to solve a problem are "embedded" in or deducible from the story of the learning anchor, and not given explicitly (as in conventional textbook problems). Supplementary data can be, as a problem component, sought out independently by the pupils (or simply given together with the problem formulation). In the example shown in Fig. 1, all the relevant data are in included in the newspaper text or are available as supplementary information in the task itself. The rationale behind this design principle is twofold: (i) it is true for problems encountered in the real world (daily life, workplaces, genuine research, etc.); (ii) the "translation" feature (OECD, 2006; cf. sect. 1) is extended by a feature of "selection" of what is relevant from what is not (for a given problem), both of which contribute to cognitive activation. Thus, "embedded data" are considered an especially important characteristic of AI (CTVG, 1991).

### 3.2 Adjustable Problem Complexity

Problem complexity is influenced both by the basic "story" (videodisc or newspaper article) and the tasks to be completed. For successful classroom application, both must be adjustable to meet the demands of a specific group of learners. This is difficult, if not impossible to achieve with video stories, but straightforward with NSP. This design principle is closely related to the concept of levels and development of competence(s). By way of example, the task variations in Fig. 3 correspond to markedly different levels of knowledge application by involving the strongly simplified case of homogenous motion (version 1) and the more challenging case of the parabolic trajectory (version 2; this latter corresponds to a typical problem in German secondary level II). This demonstrates that, by means of relatively simple variations, NSPs can be tailored to different groups of learners or competence levels and can even be used as a tool for varying task complexity to meet the demands of subgroups of students in a heterogeneous classroom.

### 3.3 Active and Autonomous Problem Solving

It is a basic tenet of physics (and general) education that successful learning requires active participation by pupils. In its original form, AI follows a pronouncedly learner-centred approach. This has been achieved with NSPs using an instructional method emphasising task-based learning (Peterßen, 2004) using an "investigation elaboration" cycle (see Fig. 4 and Schmidkunz & Lindemann, 1992). This method can be grouped within the broader class of inquiry-based learning approaches (Minner et al., 2010) and has the structure of the 1+6 steps shown in Fig. 4.

In this instructional method, the teacher introduces the pupils to the nature of the work in the investigation elaboration by presenting its different phases and the functions of each phase within the learning process.

---

[3] The question of the extent of this modification, and that of possible difference to the original AI approach which follow from it, are addressed in the following section.

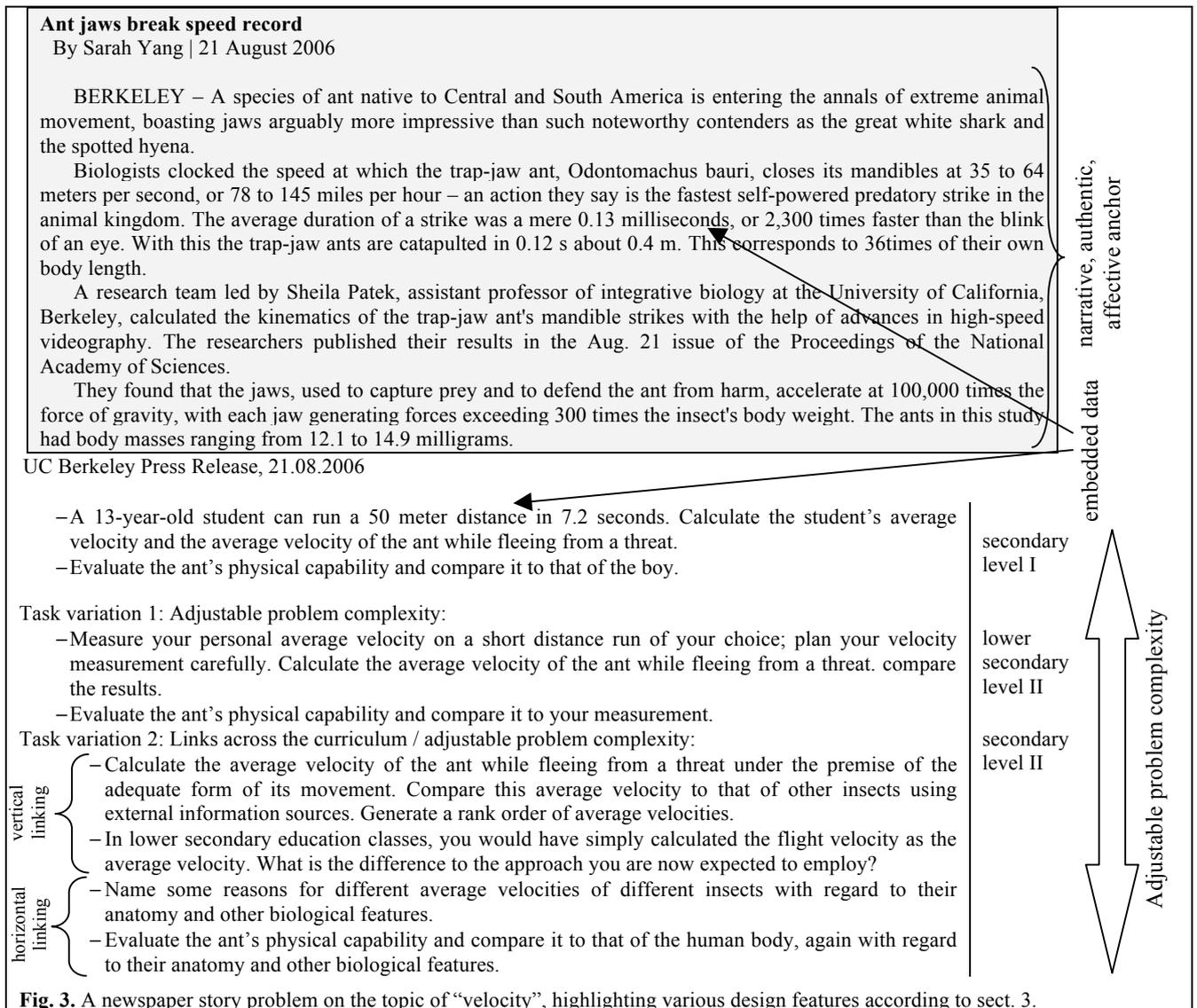

**Fig. 3.** A newspaper story problem on the topic of "velocity", highlighting various design features according to sect. 3.

Next, in the information phase, small groups of learners get the instructional materials before they start to study and discuss the newspaper clipping and the related tasks. If needed, additional information must be inferred or looked for. In the planning phase, possible approaches to solve the problem are then elaborated by identifying the necessary data (mostly "embedded", see above), and by discussing various solution approaches and their adequacy.

**Tab. 2.** Guidelines for creating newspaper articles

| No | Design principles | Guideline/Realization |
|---|---|---|
| 1 | Affective anchor media, authenticity | Newspaper articles (more than one; see No. 5) the content of which addresses the classroom topic in question |
| 2 | Story character, "narrative" format | The article should be as complete as necessary (to ensure story character), but as short as possible (to keep the complexity appropriate to the audience) |
| 3 | Embedded data | Depending on the learning goal and students' prior knowledge, the data necessary for solving the tasks should be included more or less completely in the newspaper article or in the associated tasks |
| 4 | Links across the curriculum (horizontal and vertical) | The tasks should be designed in such a way that they include interdisciplinary aspects and content across grades. |
| 5 | Related problems (by pairs or multiple) | Identify more than one newspaper articles with suitable content (cf. No. 1) and cross-link the corresponding tasks n |
| 6 | Complex problems | Design the tasks related to the newspaper articles considering different problem complexity appropriate to the audience (e.g., prior knowledge, transfer level etc.); prepare support to scaffold learning. |
| 7 | Active and autonomous problem solving | Implement NSP following classroom approaches known to support active and autonomous participation of learners (s |

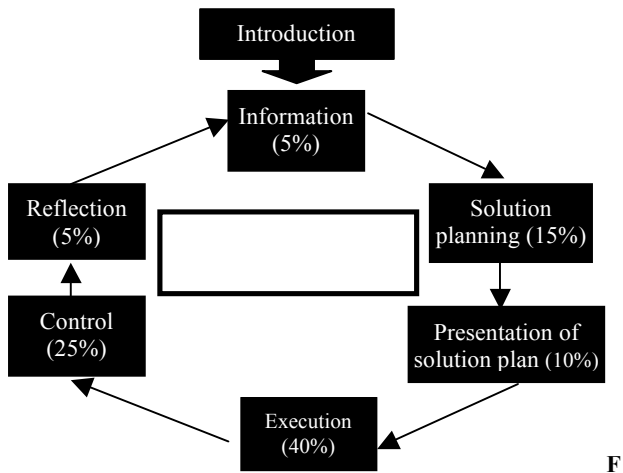

**ig. 4.** Active and autonomous learning-based tasks within an "investigation elaboration" cycle (approximate percentages of total time given for each stage given). Moderation and scaffolding are provided by the teacher throughout the cycle.

Teaching practice has shown that it is useful to have the groups present their ideas thus far to the whole class for discussion. This pre-presentation also enables groups with less well elaborated results or students with lesser domain-related abilities to benefit from their classmates' thinking. Subsequently, the problem-solving approaches elaborated thus far are worked out in the execution phase. Requiring groups to note down their reasoning and results on a single common working sheet has shown success in promoting co-operation within groups. In this way, the danger of "isolated" work that has not been adequately elaborated by the whole group can be minimized. In the subsequent control phase, groups once again present their results to the class, and their soundness and possible corrections are discussed. Last, the groups' entire solution process and results are discussed by the learners and the teacher in the reflection phase. The teacher provides the groups with individual feedback while the learners can actively discuss reasonable modifications or improvements for their approach.

Note that this specific form of active and autonomous problem solving was implemented for both the treatment and control groups, to keep all factors constant except for the one under investigation (viz., NSP versus conventional tasks).

### 3.4 Affective[4] Anchor Media, Story Character, Authenticity

Finally, "anchor media" featuring authentic, "real-life" flavoured problems to use as affectively and cognitively engaging stories, are of course essential for (M)AI. The story character or "narrative" format (as it is called sometimes) of anchor media is considered to be a particularly promising educational approach and there is a considerable body of research supporting its use in education, for motivation (Shores et al., 2009), memory (Salkind, 2008) and understanding (Mandler, 1984; Zabel, 2004); see Kuhn & Müller (2014) for a detailed discussion of these issues.

Obviously, in a first development step, both features (motivating story character, authenticity) of a given NSP will be evaluated by the designers themselves (i.e., the teacher or researcher). However, the assumption the features in question will be perceived in the same way by designers and pupils is far from certain. For the available PISA science tasks, Weiss & Müller (2015) have shown that pupils' perceived authenticity and interest are medium at best, contrary to the basic assumption of PISA, and that striking differences exist in the perception of teachers, the latter having a substantially more positive perception.

For proper control, all (M)AI learning media must thus be submitted to a test of motivation and "authenticity/real-life connection", as perceived by pupils, in a second development step. This is a critical issue for CBSE as a whole. Indeed, there otherwise persists a risk that learning problems (or more generally, any instructional material) might represent a merely "external" (i.e., for the designer) or "alleged" context, not at all perceived as such by learners (which of course is the property that really counts). In empirical research, this second step is known as a "manipulation check", in the sense of whether the feature of the learning material actually being "manipulated" (in the present case, the NSP format) in fact influences a relevant psychological variable in the desired way (in the present case, "authenticity/real-life connections"). To date, a set of more than 40 NSPs (and other types of MAI problems) has been evaluated with a positive outcome on more than 1,600 students (see below).[5]

Reviewing the above set of design principles, one sees that NSP resembles the original AI approach (CTGV, 1990; 1991) in many respects. However, it differs with respect to two specific, but important features. In contrast to AI, affective anchor media used in MAI are not limited to the video presentation format. Furthermore, "problem complexity" is not unchangeably high as in AI but can be adjusted. As explained above, both modifications serve the purpose of greater practicability and flexibility. We now turn to a discussion of the outcomes of using NSPs.

### 4. STUDY BACKGROUND AND OUTCOMES

Several studies on motivation and learning outcomes of using various MAI anchor media were carried out in different streams of the German secondary level I education system (level 2.4.4 according to the International Standard Classification of Education (ISCED) (UNESCO, 2011); cf. also Mullis et al., 2008). In this section, some basic features and the essential outcomes of these studies are summarised (for more details, cf. Kuhn (2010), Kuhn & Müller (2014)).

---

[4] "Affective" is a very broad term and is used, among other things, as an umbrella term for emotional and motivational dimensions including interest and self-believes (Alsop & Watts, 2003)

[5] This does not yet include ongoing research on "comic strip" and similar kinds of problems, see sect. 5.

## 4.1. Study Background

The investigation took place within a co-operation network (see Tab. 3) comprising physics teachers and a research group based at a university physics teacher education unit. The sample encompasses over 1,600 pupils (treatment group: ca. 785; control group: ca. 815) in a variety of teaching settings, notably in schools with differing academic levels (7th/8th graders: average age 12.8 years; 56% female; 44% male; 9th/10th graders: average age 16.3 years; 52% female; 48% male). The topic taught was "velocity" (elementary kinematics) in 7th/8th grade, and "electrical energy" in 9th/10th grade; see figures 1, 3 and 5. For the advertisement tasks (see sect. 5), the topics were "thermal capacity" and "caloric value" (Vogt, 2010).

Studies were carried out within regular classroom teaching as a comparison of intervention ("treatment") and "control" classes (EC and CC, respectively).[6] In the experimental condition, classes worked with newspaper story problems (see Fig. 1a and 5a), while classes in the "control condition" had to work on "conventional" textbook type tasks (see Fig. 1b and 5b). The instruction followed the time course presented in Tab. 4.

The two groups worked on different worksheets with tasks dealing either with electrical energy or kinematics. Learning content and difficulty in the two conditions were identical. The newspaper story problems in the EC differed from the tasks in the CC only in layout and language style (newspaper vs. textbook) but were identical in terms of basic information and related questions (see Fig. 1c and 5c).

The procedures concerning worksheets and tests in EC and CC were identical, too: in each lesson, the pupils received the instructional material and worked on it following the investigation elaboration cycle shown in Fig. 4.

Worksheets consisted of learning (or "exercise") tasks for basic understanding as well as transfer tasks. The treatment of each worksheet lasted two 45-minute school lessons, and the duration of the complete instructional unit was three weeks with one or two physics lessons per week (in compliance with curriculum regulations).

Selection and evaluation (comprehensibility, difficulty level) of the tasks for the learning worksheets and for assessment were carried out by the abovementioned physics education co-operation network. Problem difficulty was operationalised according to the PISA proficiency levels, including transfer tasks[7] (see Bybee, 1997 and Baumert et al., 2002; it is understood that a given problem difficulty level requires the same learner's proficiency level $x$ to be solved). Only problems with satisfying interrater agreement for curriculum validity and difficulty level were retained (according to standard methods of quantitative educational research; cf. Kuhn (2010)). This includes whether a problem involved transfer of knowledge, one of the main issues of the present work (cf. Introduction).

To minimise teacher effects, each participating teacher taught an EC and a CC class. Other control measures took account of possible differences in teacher engagement or pupil ability in the two different groups, novelty effects, and other possible influences. These are described in detail in Author (2010) and Author (2010).

**Tab. 3.** LeBiNet – a research-practice co-operation network for physics education (Kuhn et al., 2008)

| N pupils | > 1,600 |
|---|---|
| N classes | > 40 |
| N teachers | > 20* |
| N schools | 15 |
| N school types | 6** |

*exact numbers depend weakly on the details of investigation considered (e.g. final dates of data collection)
**this covers all the ordinary school types of secondary level I of the federal state where the studies took place: *Hauptschule* (basic general education); *Realschule* (extended general education), *Gymnasium* (advanced general education) and combined/integrated forms of these (see Mullis et al., 2008).

## 4.2. Motivation and Learning Outcomes

Within the studies, pupils' motivation and learning achievement were measured before (motivation only), immediately after, and up to two months after treatment. Where available, standardised tests were used. In some cases, however, tests for the specific demands of this research had to be developed.

For the "electrical energy" topic, motivation and achievement results are presented in figure 6. Percentages refer to the maximum value of the scale of interest (percentage of maximum possible, POMP; Cohen et al., 1999). While motivation (Fig. 6a) was nearly equal in EC and CC before treatment, pupils in EC reported significantly higher motivation than those in CC during and after treatment, as well as in follow-up measures. Percentage gains of EC compared to CC are more than 20% for post- and almost 20% for the follow-up test, respectively (see Tab. 5a).

Thus, the newspaper story problem task format indeed fosters student motivation. Moreover, this effect is still apparent two months after treatment, indicating that the enhancement of student motivation by newspaper story problems is sustainable at least in the medium term.

---

[6] A "quasi-experimental" situation, in contrast to laboratory conditions.
[7] The various conceptualisations of the PISA competence levels include both near and far transfer (e.g. levels III and IV according to Baumert et al., 2002). In the limited space of this contribution this differentiation is not taken into account, see however Kuhn (2010).

**Tab. 4.** Time course of instruction in the two comparison groups

| Week | Control group | Treatment group |
|---|---|---|
| 1 | Tests of non-verbal intelligence and reading comprehension | |
| 1 | Motivation pre-test (MOT1-PRE) | |
| 4 | Conventional problems Worksheets 1–3 | Newspaper story problems Worksheets 1–3 |
| 5 | Achievement test | |
| 5 | Immediate motivation test (MOT2-POST) | |
| 6…13 | Conventional instruction of a new topic | |
| 14 | Follow-up motivation test (MOT3-FUP) | |

| (a) "Newspaper story problem" format | (b) "Conventional problem" format |
|---|---|
| **GLIDER FLIGHT OVER THE CHANNEL** **An Austrian extreme athlete** has crossed the channel with a special paraglider. After leaping from a plane at a 9000-metre height, Felix Baumgartner glided over the Channel with an initial speed of approx. 300 kilometres per hour yesterday morning. He covered the 34-kilometre distance between Dover and Calais in 14 minutes. The 34-year-old car mechanic landed safely in Calais at 6.23 a.m. "I felt like a bird," the extreme athlete enthused after his flight. "In spite of some problems with the ropes and a torn canvas, it was a fantastic experience." The Austrian had prepared for the adventure over a period of three years. Baumgarter had previously jumped from the 451-metre high Petronas Towers in Kuala Lumpur and completed the world's shortest parachute jump from the statue of Christ in Rio de Janeiro (ap/rtr/phote:ap) *DIE RHEINPFALZ*, 01.08.2003 | A man into doing extreme sports jumped with a paraglider from a plane in 9000 m height above sea level. He glided with an initial velocity of about 300 kilometers per hour. It took him 14 minutes to cover the 34 kilometer distance between Dover and Calais. |
| **c) Questions common to both formats** 1. What was the glider's average velocity on his channel crossing its flight across the channel? 2. Compare your answer to No. 1 to the velocity given by the text. Why are the velocities different? | |

**Fig. 5.** A newspaper story problem (a), and its conventional counterpart (b), topic "velocity". Tasks are identical in both cases (c).

The key question, however, is whether this enhancement of motivation is accompanied by an effect on learning achievement. Indeed, the latter is often missing in research settings (see e.g., Bennett et al., 2007) and in practice, even when motivational effects are present.

In our research, considerable differences in achievement between EC and CC were found (see Fig. 6b). Percentage gains of EC compared to CC were more than 15% for the post-test and almost 30% for the follow-up test, see Tab. 5b.

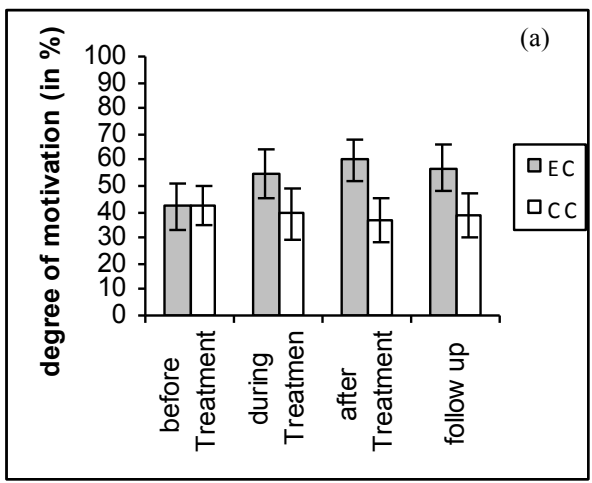
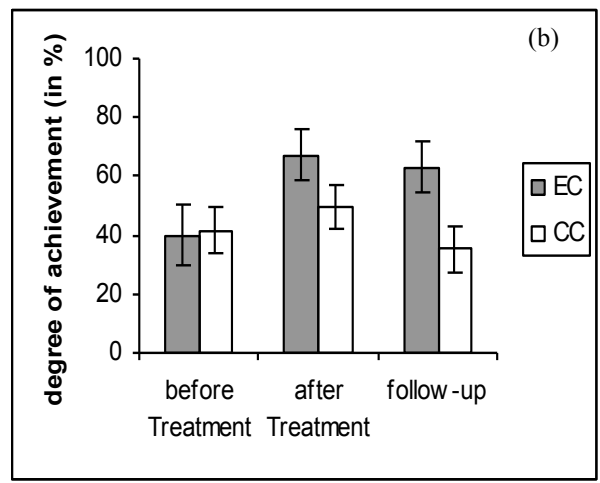

**Fig. 6.** Motivation (a) and learning achievement (b) over time: average values and standard deviations (error bars) in EC (newspaper story problems, grey) and CC (conventional tasks, white) for the subject "electrical energy"; y-axis: percentage of maximum possible (for more details see Kuhn (2010, p. 173 ff)). Note that the value "before treatment" refers to the grade average (included to show comparability of CG and EG), and *not* to the specific knowledge test about the teaching unit.

The findings reported thus far are related to motivation and learning in general. There are, however, specific aspects of particular importance for the present study, namely. authenticity (as a component of motivation) and transfer (as a component of learning), to which we now turn. Regarding authenticity, there is strong evidence that pupils indeed perceive NSPs as "authentic" (Fig. 7), avoiding the potential pitfall associated with merely "external" contexts; as mentioned above, this is an essential issue for all CBSE (see sect. 3, features "authenticity/affective anchor media"). As for transfer, the NSP classes also showed a considerable increase (Fig. 8). Percentage gains of EC compared to CC are almost 15% for the post- and more than 25% for the follow-up test, respectively. Thus, a core learning objective of CBSE, namely, to overcome inert knowledge (cf. Introduction), was achieved using the present approach.

A final set of findings with high relevance for both classroom practice and research concerns possible influence factors (in the sense of covariates): for all positive effects on motivation and achievement, no influences (or only weak ones) were found for gender, reading competence, physics, mathematics, or German language grades. Nor were such influences found for features of the broad variety of teaching settings where NSPs were tested, such as the teaching style of different teachers or the academic level of the different school types involved in the study (cf. 4.1). Note that this independence holds for gains in motivation and learning (the absolute values depend e.g. on prior knowledge). This independence from grades (individual level) and school type (institutional level) as indicators of general educational background is of particular interest, as several sophisticated educational strategies have repeatedly been shown to have benefits for students at a high educational level only (cf. "Matthew effect", Hattie, 2009). Of course, absolute achievement level does depend on these factors, but when comparing EC and CC, students of "various backgrounds" and educational levels are likely to profit from the NSP format to the same extent. Again, more details on methodology for these (and other) control aspects of the studies can be found in Kuhn (2010), and Kuhn & Müller (2014).

**Tab. 5.** Descriptive data of (a) motivation and (b) achievement (means (M) and standard deviations (SD); percentages are POMP values as above) for the topic "electrical energy" in the different classes.

| (a) Motivation data (in %) | | | | (b) Achievement data (in %) | | |
|---|---|---|---|---|---|---|
| Motivation… | EC M(SD) | CC M(SD) | | Achievement… | EC M(SD) | CC M(SD) |
| … before treatment | 42.1 (17.3) | 42.7 (14.8) | | … before treatment | 39.9 (20.4) | 41.7 (15.3) |
| … during treatment | 54.7 (18.9) | 39.3 (19.4) | | … during treatment | n/a | n/a |
| … after treatment | 60.1 (16.4) | 36.5 (16.8) | | … after treatment | 67.2 (17.0) | 49.7 (15.5) |
| … follow-up | 57.0 (17.7) | 38.7 (17.6) | | … follow-up | 63.0 (17.5) | 35.4 (15.7) |

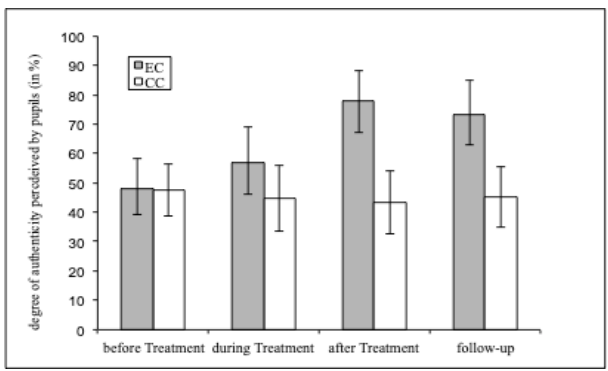

**Fig. 7.** Authenticity of the tasks, as perceived by pupils.

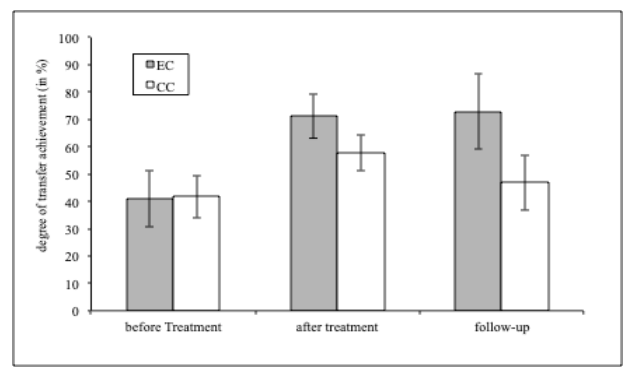

**Fig. 8.** Achievement on transfer tasks for the topic

**For both figures**: y-axes = POMP (Cohen et al., 1999); grey: EC average, white: CC average; error bars: standard deviations (*not* standard error of averages); for more details see Kuhn (2010). Value "before treatment" of Fig. 8: see Fig. 6.

## 5. CONCLUSIONS AND PERSPECTIVES

The present article reports a large-sample, quasi-experimental intervention study using "newspaper story problems" (NSPs) as a specific form of context-based science education (CBSE). On the one hand, there is ample support for such an approach from good practice in many science (and non-science) classrooms. On the other hand, it is based on extensive educational science research on "story contexts", as well as on a well-studied instructional design, "Anchored Instruction" (AI), which features detailed design principles for learning media in such contexts ("anchor media"). The guiding idea of our study is to use this valuable research background and to develop and investigate forms of applications useful for a variety of practical classroom conditions. It is organised in the framework of a co-operation network of physics teachers and a university physics education research and teacher education unit (Müller et al., 2009). This allows for a large sample and variety of pupils, teachers, and schools, and ensures the intended interaction between research and practice takes place in a close, collaborative manner throughout the investigation).

The results show large positive effects on pupils' motivation and learning; in terms of effect sizes[8] (Cohen *d*), the advantage of NSPs is > 1.5 for motivation and > 0.9 for learning (see Kuhn, 2010 and Kuhn & Müler, 2014 for details). This holds also in particular for perceived authenticity and for transfer. Thus, two basic issues for the whole CBSE school of thought, that is to say, perceived (as opposed to merely "outside" or "assumed") authenticity as an essential motivation factor and transfer as a central learning objective are overcome with the present approach. Moreover, all positive effects (especially transfer ability) are stable for several months at least. These positive effects are also "robust" in the sense of being explicitly not (or only weakly) affected by possible influence factors on the individual (gender, various grades, prior motivation, and prior knowledge) or classroom/school (various teachers, schools, and school types) level.

A potential difficulty of the approach might be a dependence on reading ability, for which however no evidence was found. One could assume that the effects depend on the topicality of the newspaper articles, in the sense that the more topical the greater the motivation and learning effect on learners. However, our results show that the instructional materials used lead to positive effects even when the articles used were several years old. The practical aspects of its usability and adaptability to content, difficulty and competence level, as well as its embedding in a broader task culture, also make the medium very interesting for teachers. Several aspects not considered in the study, but important for today's society, such as critical reasoning or data literacy, can be easily added. Finally, with rapidly changing media and media use, other sources than printed newspapers would be interesting to test.

Note also that within the instructional framework of the present contribution (sect. 3.3), the intervention took place in the "information" part of the tasks, not the questions actually to be resolved, nor the support, nor the ensuing steps of the approach (Fig. 4). This was a deliberate limitation of the present work, first for methodological reasons (vary only one/few factors at a time), second for reasons of feasibility and compatibility (not requiring far-reaching changes of the overall instructional framework). As the empirical data shows, also this limited approach has encouraging results, yet combining it with changes of other elements of the instructional framework are is course possible, and maybe interesting for future research.

In summary, the use of NSPs has large, positive, robust, and sustainable effects on both motivation and learning. Together with its practicability and flexibility, this implies a desirable degree of "classroom usability". That is, intermediate-scale CBSE approaches (in the sense of sect. 3) using anchor media that can be designed with relatively little effort and resources, have been shown to be an effective element or "building block" for science education.

The following comment is in order. For the reasons mentioned in section 3, the present approach differs from AI by the extent or scale to which the classroom setting must be modified. This is an important issue for many instructional approaches in general, and CBSE in particular.

For the latter, examples range from "micro-contexts" to "macro-contexts", or (in the terms of Bennett et al. (2007)) from "enrichment materials" (with a duration of some weeks) to "full courses" (duration around one year and longer). A difference in scale of a given intervention is frequently associated with a difference in the extent to which it introduces changes to conventional teaching practices. Note that these differences of extent exist on a continuum containing intermediate forms, such as, for example, a brief introduction of background knowledge followed by extended learning phases involving motivating, real-life applications and largely based on active, autonomous investigation by the pupils. With respect to the "teaching script", the NSP intervention presented here is of this latter, intermediate type (cf. figure 4). It is here that it differs most from the original AI, "macro"-type approach, despite the many design principles inspired by and shared with the latter.

But we see plausible and strong arguments to investigate NSPs as an example of an intermediate scale change within the CBSE philosophy. They are based on an inexpensive and easy-to-obtain type of media (compared to multimedia videodiscs), they are compatible with a wide variety of teaching styles and classroom conditions, and the instructional features of their usage (from problem complexity to classroom implementation) can be varied to a large extent. Note, moreover, that NSPs are in no way opposed to macro-context approaches. Rather, we see them as a way to obtain elements or "context-building stones" which, if successful, allow a more detailed understanding and application of different macro approaches.

---

[8] The basic definition of an effect size is $(M_T - M_C)/SD$, where $M_T$ and $M_C$ are the means (of some variable of interest) for the treatment and control groups, respectively, and $SD$ is the either the pooled standard deviation or that of the control group (Cohen, 1988). In simple terms, $d$ thus measures the impact of an intervention in units of standard deviations of the sample under consideration. Usual effect size levels (as established from comparison of a great many of studies in different areas, Cohen (1988)) are small ($0.2 < d < 0.5$), medium ($0.5 \leq d < 0.8$) or large ($0.8 \leq d$). Hattie (2009) uses a threshold level ("hinge point") for practical significance of $d = 0.4$.

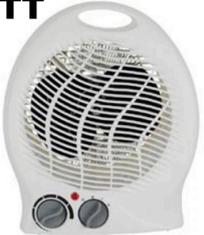

**Fig. 9.** An "advertisement problem" with original ad clipping (left), and problem questions (right). Note again, that problem complexity can be varied over a large range.

On a cool morning, you would like to heat your bathroom ($A = 20$ m$^2$) from 12 °C up to 20 °C with a heater (see left).
1. What time does it need to warm up your bathroom?
2. What is the costs for heating the bathroom, given a price of 30 cent per kilowatt hour?
Note: Assume that the electrical energy is converted completely into heat energy of air.

It is worth noting that there is a recent discussion on different aspects and conceptualisations of CBSE which we want to mention as a broader theoretical perspective (Löffler & Kauertz, 2021; Broman et al., 2020; Löffler et al., 2018; Bellocchi et al. 2016; Prins et al. 2016); however, within the limited focus of this research-based report on practice, we do not discuss this further here.

Finally, the framework for classroom implementation (cf. sect. 3) and the co-operation network (cf. sect. 4.1 and Tab. 2) in this study offers a useful starting point for research on further variants and more detailed versions of relevant hypotheses. We now provide an outlook on some of these.

First, similarly to newspaper story problems, advertisements with some physics (or more generally, science-related) content can be used to provide learning material and opportunities with a focus on real-world connections. An example of such "advertisement problems" is shown in Fig. 9 and a detailed investigation of this type has already been undertaken (Vogt, 2010). Another example are comic strip problems, which are, again similarly to NSPs, based on some comic strip sequence featuring science-related issues and assking pupils to work on various questions and related tasks (Kuhn et al. 2010b).

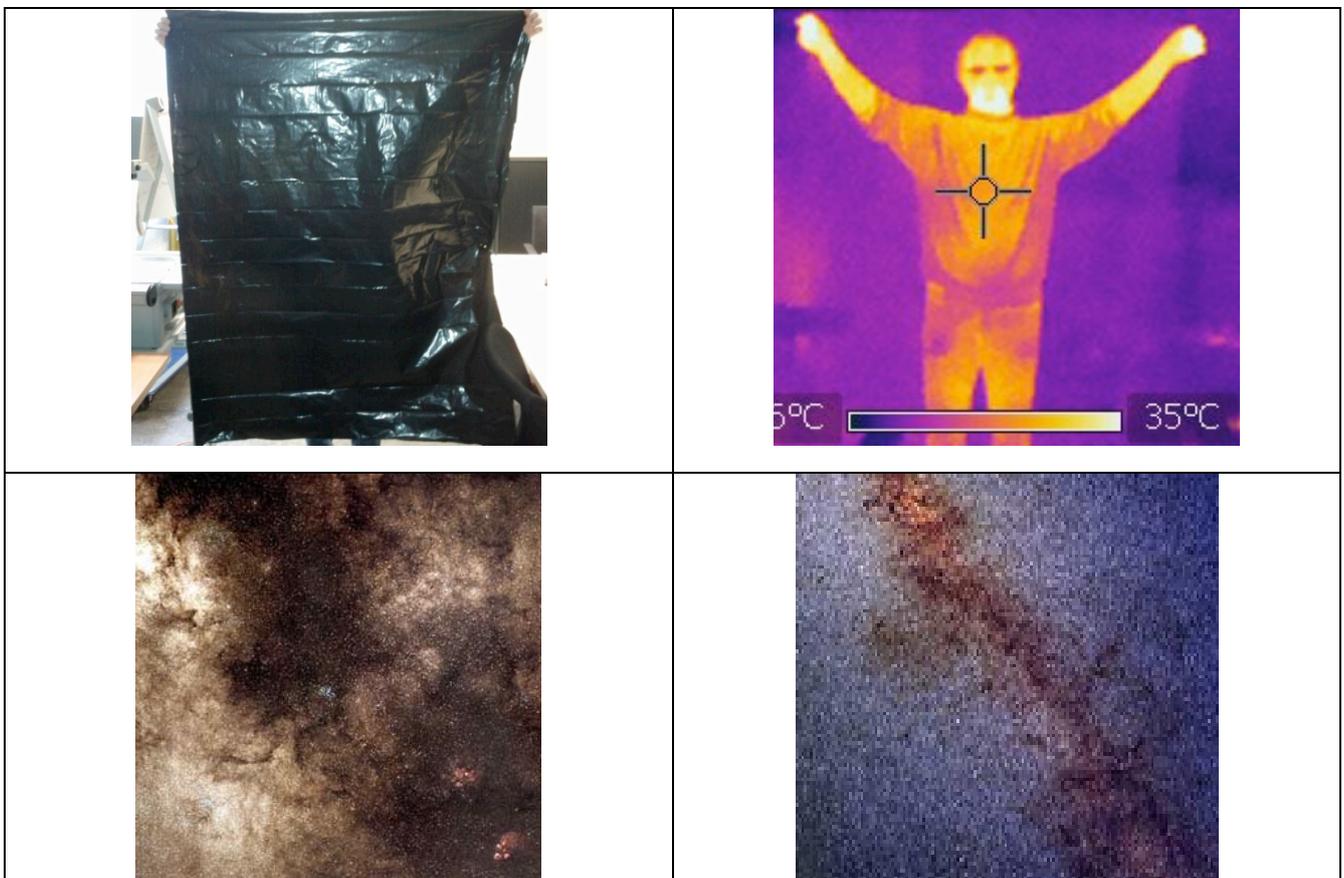

**Fig. 10.** Looking behind the curtain: Infrared images and selective absorption from human to Milky Way. Top: Behind a plastic "curtain" (polyethylene): person in visual range invisible (left), in infrared range visible (right). Bottom: Behind the dust "curtain" (interstellar dust): galactic centre in visual range invisible (left), in infrared range visible (right)

Second, a form of experimental context-based physics learning has been developed and investigated in a similar framework in the last decade, with a broad range of activities using smartphones and tablet computers as portable physics sensors (Kuhn & Vogt, 2012; Vieyra et al. (2015); Darmendrail et al., 2019). The idea is to establish real-life connections through a material context, using familiar devices that already play a large role in the everyday life of youths, but in an unexpected, and hopefully interesting way (see Hochberg et al. (2018) for an empirical study) and with acceptable learning effects (Becker et al., 2020a; 2020b; Hochberg et al., 2020). This idea has seen strong development internationally (Kuhn & Vogt, 2012; Gil & Di Laccio, 2017; Kuhn & Vogt, 2022), and in a broad range of applications in the physics education groups of the authors (Klein et al, 2014; Kuhn et al., 2015; Müller et al., 2016; Keller et al., 2019; Darmendrail & Müller, 2020).

Third, visual (aesthetic or fascinating) contexts for science learning and their potential for providing motivation and cognitive activation are investigated in another strand of research development (Müller et al., 2012; Lenzner et al., 2013; Lamparter & Girwidz, 2020); see Fig. 10 for an example related to astronomy (Koupilova et al., 2015).

Research on these different forms of context-based tasks, and on various other questions of interest (such as the necessary "dose" (duration) of the intervention required to achieve a desired educational effect) is underway, always with a close eye on classroom application. On all these and other, potentially related issues, the authors are open to discussion with *PriSE* readers, especially about possible co-operation (quantitative evaluation, comparison of data, etc.) with teachers who use CBSE approaches similar to those described in the present work, or who would be interested in doing so.

## ACKNOWLEDGEMENTS

Financial support by the Ingrid-and-Wilfried-Kuhn-Foundation for Physics Education is gratefully acknowledged, as are valuable discussions with Prof. Dr. Wieland Müller, and support from Alex Brown (St-Genis-Pouilly, France) for the preparation of the manuscript.